# Understanding the Memory Window Overestimation of 2D Materials Based Floating Gate Type Memory Devices by Measuring Floating Gate Voltage

*Taro Sasaki, Keiji Ueno, Takashi Taniguchi, Kenji Watanabe, Tomonori Nishimura, and Kosuke Nagashio\**

T. Sasaki, Dr. T. Nishimura, Prof. K. Nagashio
Department of Materials Engineering
The University of Tokyo
Tokyo, 113-8656, Japan
*E-mail: nagashio@material.t.u-tokyo.ac.jp

Prof. K. Ueno
Department of Chemistry
Saitama University
Saitama, 338-8570, Japan

Dr. T. Taniguchi
International Center for Materials Nanoarchitectonics
National Institute for Materials Science
Ibaraki, 305-0044, Japan

Dr. K. Watanabe
Research Center for Functional Materials
National Institute for Materials Science
Ibaraki, 305-0044, Japan

## Abstract

The memory window of floating gate (FG) type non-volatile memory (NVM) devices is a fundamental figure of merit used not only to evaluate the performance, such as retention and endurance, but also to discuss the feasibility of advanced functional memory devices. However, the memory window of two dimensional (2D) materials based NVM devices is historically determined from round sweep transfer curves, while that of conventional Si NVM devices is determined from high and low threshold voltages ($V_{th}s$), which are measured by single sweep transfer curves. Here, it is elucidated that the memory window of 2D NVM devices determined from round sweep transfer curves is overestimated compared with that determined from single sweep transfer curves. The floating gate voltage measurement proposed in this study clarifies that the $V_{th}s$ in round sweep are controlled not only by the number of charges stored in floating gate but also by capacitive coupling between floating gate and back gate. The present finding on the overestimation of memory window enables to appropriately evaluate the potential of 2D NVM devices.
(173 words < 200 words)

In recent years, two-dimensional (2D) materials including semimetallic graphene, semiconducting transition metal dichalcogenides (TMDs) and insulating *h*-BN have emerged as potential next generation electronic materials. For instance, semiconducting TMDs have been studied as channels in field effect transistors (FETs), since their atomically thin nature can suppress the short channel effects that scaled Si-FETs have faced.[1-3] Not only *n*-MoS$_2$ FETs[4,5] but also ambipolar WSe$_2$ or MoTe$_2$ FETs[6-8] have been widely investigated from the context of channel mobility, contact engineering, etc. In addition, to explore their intrinsic properties due to the dangling bond free interface, 2D hetero-stack technology enables the fabrication of more advanced device structure[9] where insulating *h*-BN is applied as the atomically flat substrate[10,11] and gate insulator[12,13] with SiO$_2$ compatible permittivity (2~4).[14] As well as FETs, 2D materials have been utilized in non-volatile memory (NVM) devices, which are one of key building blocks for modern integrated circuits.[15,16]

In particular, lots of efforts have been devoted to floating gate (FG) type memory devices. Their device structures have also evolved from the early stage of $MoS_2$ channel/$HfO_2$ tunnel barrier/multilayer graphene FG[17] stacks to recent 2D hetero-stack of $WS_2$ channel/$h$-BN tunnel barrier/graphite FG,[18] where the large memory window over 20 V for gate voltage ($V_G$) range of $\pm 25$ V and a good retention characteristic of 13% charge loss after 10 years have been demonstrated. Interestingly, many other 2D NVM devices, such as black phosphorus channel device and so on,[19-22] have also exhibited the large memory window, regardless of material systems, even though the origin has rarely been discussed. Moreover, 2D hetero-stacks have recently been further extended to artificial synaptic emulators for neuromorphic computing.[23]

As mentioned above, the memory window, which is defined by the threshold voltage ($V_{th}$) difference between program state and erase state, is a fundamental figure of merit for FG type NVM devices.[24] Although ON/OFF current levels are readout in practical operations, the evolution of $V_{th}$ difference against time or cycles is quite important to investigate the physical mechanisms of retention and endurance since $V_{th}$ shift indicates the polarity and the number of ejected carriers.[25,26] Furthermore, memory window is also used to discuss the feasibility of advanced functional memory devices, such as multi-bit-cells and synaptic devices. In the 2D based memory research field,[17-23] the memory window has historically been determined from $V_G$ round sweep transfer curves from the pioneer work on graphene NVM device,[27] which is similar to the hysteresis measurement. In the Si based memory research field,[26,28,29] on the other hand, the memory window is determined from high and low $V_{th}$s, which are measured after program or erase (P/E) operation. That is, the P/E operation is conducted by applying the pulse voltage to the control gate. After the P/E operation, single sweep drain current ($I_d$) - $V_G$ measurement performed with a small $V_G$ range determines both high and low $V_{th}$s. Therefore, it can be said that memory window determination methods strongly depend on the research fields. However, there are no discussions on which method should be used to determine the memory window of 2D NVM devices. This is critical to fairly evaluate the potential of 2D NVM device applications, which is inherently higher than that of conventional Si NVM technologies.

In this study, memory windows of fabricated 2D NVM devices were determined by the aforementioned two methods. Indeed, the memory window determined from round sweep transfer curves was overestimated compared with that determined from $V_{th}$ measurements after the P/E operation. The floating gate voltage ($V_{FG}$) measurement proposed here revealed that the origin for this discrepancy results from the fact that $V_{th}$s in round sweep are controlled not only by the number of charges stored in FG but also by capacitive coupling between FG and back gate (BG). Therefore, in case of 2D NVM devices, the memory window should be determined from $V_{th}$s measured by single sweep after the P/E operation. This understanding enables the discussion of the performance, such as retention and endurance, in the same context as the Si NVM devices.

Schematic and optical microscope images of fabricated 2D based memory device are shown in **Figures 1a** and **1b**, respectively. It consists of a 2H-$MoTe_2$ channel (8.3 nm), an $h$-BN tunnel barrier (15 nm) and a graphite FG (6.9 nm) on the global back gate of a 90 nm $SiO_2/n^+$-Si substrate. Ni and Au were used as metal electrodes. Since tunneling across the $h$-BN tunnel barrier is crucial to determine the memory window, ambipolar $MoTe_2$ was selected as the channel material to ensure both electron and hole tunneling. Note that the carrier polarity for tunneling in graphene/$h$-BN/metal system has already been revealed to be positive (hole).[30] These 2D materials are stacked sequentially by a dry transfer system[31] with the help of an aluminum slope to suppress bubble formation at the 2D heterointerface, as explained in **Figure S1**.[32] The atomic force microscope (AFM) image and Raman spectra of typical $MoTe_2$/$h$-BN/graphite hetero-stack based memory device are shown in **Figures S2** and **S3**, respectively. In most previous studies, since source and drain metal electrodes overlapped the FG through the tunnel barrier, two different tunneling paths, that is, channel to FG and metal to FG should be considered. To ensure the single tunneling path of channel to FG, the access region was designed (no overlapping between the channel region and FG), as shown in **Figure 1b**.

Three kinds of measured $I_d$ - back gate voltage ($V_{BG}$) characteristics with different sweep ranges and sweep directions are shown in **Figure 1c**. The open and filled circles represent the start and end points in the $V_{BG}$ sweep. The black curve indicates the round sweep $I_d$-$V_{BG}$ for the $V_{BG}$ range of $\pm 30$ V. The ambipolarity of $MoTe_2$ and the large memory window are clearly observed, which is consistent with the typical memory operation of 2D NVM devices reported previously. In this measurement, the potential of the FG was floating. When the FG was grounded during the round $V_{BG}$ sweep, the large memory window disappeared, and quite a small hysteresis resulted, as shown in **Figure S4**. This suggests that the graphite FG acts as the charge trapping layer and could contribute to the large memory window when its potential is floating. Interestingly, the slope of $I_d$ against $V_{BG}$ is gradual in positive sweep, while that is steep in negative sweep. Since the $V_{th}$ shift due to the charge stored in the FG by tunneling between the channel and FG does not change the slope in the $I_d$-$V_{BG}$ curve, the different slopes suggest the existence of a factor other than the $V_{th}$ shift, which will be discussed later. On the other hand, the red (blue) curve indicates single sweep $I_d$-$V_{BG}$ measured from $V_{BG} = 0$ V after the program (erase) operation, which was conducted by applying +30 V (-30 V) to the BG for 10 s for the complete operation. Although 10 s programing is quite long for typical NVM devices, the operation speed is not the main point of this study. Here, to compare the memory windows determined by the two different methods, $V_{th}$ was defined as $V_{BG}$ at $I_d = 200$ pA for simplicity. It is evident that the memory window of ~30 V extracted from the round sweep is much larger than that of 11.6 V extracted from the difference between the two single sweeps.

To clarify the reason for the memory window difference, a more detailed understanding of the operation mechanism behind the $I_d$-$V_{BG}$ curve is required. Therefore, the floating gate voltage ($V_{FG}$) measurement in the $I_d$-$V_{BG}$ round sweep is proposed here, because $V_{FG}$ is a key parameter that determines whether or not tunneling occurs between MoTe$_2$ and the graphite FG through $h$-BN. Under the condition that the additional Ni/Au electrode connected to the graphite FG was set to zero current (**Figure S5**), the trajectory of $V_{FG}$ was traced in the $I_d$-$V_{BG}$ round sweep, as shown in **Figure 2a**. Note that the $V_{FG}$ measurement itself has a negligible effect on the $I_d$-$V_{BG}$ curve, as shown in **Figure S6**. At the initial stage of the positive $V_{BG}$ sweep, $V_{FG}$ increased linearly with increasing $V_{BG}$ and then, saturated at 11.5 V. The situation of the negative $V_{BG}$ sweep was similar to that of the positive sweep. $V_{FG}$ decreased linearly with decreasing $V_{BG}$ and saturated at -9.2 V. Here, to understand the trajectory of $V_{FG}$, the tunneling current across $h$-BN was measured by applying a bias to the FG with the grounded source and constant $V_{BG}$. **Figures 2b** and **2c** show the tunneling currents for positive and negative $V_{FG}$ sweeps at $V_{BG} = 0$ V, respectively. It is evident that the two kinds of saturated $V_{FG}$ values in **Figure 2a** correspond to the tunnel start voltages for positive and negative $V_{FG}$ sweeps shown in **Figures 2b** and **2c**, indicating that tunneling takes place at the constant $V_{FG}$ region. In addition, the polarity of the tunneling carrier for both $V_{FG}$ sweep directions has been confirmed to be positive (hole) by analyzing the tunneling current modulated with various $V_{BG}$s, which is similar to the previous study.[30]

The band diagrams used to explain the $V_{FG}$ change in the positive $V_{BG}$ sweep are schematically illustrated in **Figure 3a**. (1) At the initial stage of the positive $V_{BG}$ sweep, $V_{FG}$ increases with increasing $V_{BG}$ because of capacitive coupling between the FG and BG. Tunneling between the channel and FG does not occur, because the electric field across $h$-BN is not large enough to tunnel. In this capacitive coupling region, the relationship between $V_{FG}$ and $V_{BG}$ can be described as follows,[24]

$$\frac{dV_{FG}}{dV_{BG}} = \frac{C_{ox}}{C_{ox} + C_{BN}}, \cdots (1)$$

where $C_{ox}$ is the capacitance between the FG and BG (the insulator is SiO$_2$), and $C_{BN}$ is the capacitance between MoTe$_2$ and the FG (the insulator is $h$-BN). In this device structure, the SiO$_2$/$n^+$-Si substrate acts as the global back gate, and the graphite FG is extended by an additional metal electrode to measure the potential of the FG. On the other hand, the MoTe$_2$ channel overlaps with the graphite FG with only a small area. This means that the effective area in $C_{ox}$ is considerably larger than that in $C_{BN}$. Therefore, $C_{ox} \gg C_{BN}$ leads to $dV_{FG}/dV_{BG} \sim 1$, which agrees well with the slope shown in **Figure 2a**. (2) Once $V_{FG}$ reaches the tunnel start voltage in the positive $V_{BG}$ sweep, that is, 11.5 V in this device, tunneling from the FG to the channel will start. When $V_{BG}$ further increase by a small amount of $\Delta V_{BG}$, $V_{FG}$ also increases with increasing $V_{BG}$ because of capacitive coupling. However, hole tunneling from the FG to the channel brings the potential of the FG back to the tunnel start voltage. This feedback mechanism continues as long as $V_{BG}$ is swept toward the positive direction. As a result, $V_{FG}$ is fixed at the tunnel start voltage, while $V_{BG}$ continues to increase. Similarly, the $V_{FG}$ trajectory in negative $V_{BG}$ sweep can be understood in a manner similar to that in positive $V_{BG}$ sweep. In short, the increase/decrease in $V_{FG}$ at stages of (1) and (3) is controlled by capacitive coupling, while $V_{FG}$ at stages of (2) and (4) is fixed by the feedback mechanism between capacitive coupling and tunneling between the channel and FG.

Moreover, the $V_{FG}$ trajectory also clarified the reason for the different slopes in the round sweep $I_d$-$V_{BG}$ curves. **Figure 3b** schematically explains two kinds of modes in the $I_d$-$V_{BG}$ curve. Although only the $n$-type region is considered here for simplicity, the discussion is also applicable to the $p$-type region. At first, the channel is divided into three regions, that is, two access regions and one FG region (the channel over the FG), as shown in the inset of **Figure 3b**. Here, the FG can be considered as the local back gate with the $h$-BN gate dielectric as well as the charge trapping layer. The conductance at the two access regions ($G_{access}$) and FG region ($G_{FG}$) is modulated by the BG and the FG, respectively. Therefore, the conductance of the whole channel between the source and drain is governed by the smaller of $G_{access}$ and $G_{FG}$. The FG possesses a stronger gate controllability than the BG, since the FG is located closer to the channel than the BG. As a result, two kinds of modes in the $I_d$-$V_{BG}$ curve can be considered. One mode is the FG controlled mode, in which the channel conductance is mainly modulated by the FG under the condition of $G_{FG} < G_{access}$. The other mode is the BG controlled mode, in which the channel conductance is mainly modulated by the BG under the condition of $G_{access} < G_{FG}$. In case of the positive $V_{BG}$ sweep from -30 V, $V_{FG}$ increases with increasing $V_{BG}$ due to capacitive coupling, resulting in the condition of $G_{access} < G_{FG}$. Therefore, $I_d$-$V_{BG}$ at the subthreshold region in the positive $V_{BG}$ sweep can be considered as the BG controlled mode, with a gradual increase in $I_d$ due to the smaller $C_{ox}$. With a further increase in $V_{BG}$, $V_{FG}$ saturates at the positive tunnel start voltage. Even after $V_{FG}$ saturation, $V_{BG}$ still continues to increase up to +30 V, as schematically shown by (iii) in **Figure 3a**, and $I_d$ also continuously increases with increasing $V_{BG}$. Therefore, the whole $n$-type branch of $I_d$-$V_{BG}$ in the positive $V_{BG}$ sweep can be considered as the BG controlled mode. When the negative $V_{BG}$ sweep starts from $V_{BG}$ = +30 V, $V_{FG}$ immediately decreases with decreasing $V_{BG}$ due to capacitive coupling, because hole tunneling terminates at the reverse $V_{BG}$ sweep. This leads to the steep decrease in $I_d$ at the subthreshold region in the negative $V_{BG}$ sweep, since the condition is $G_{FG} < G_{access}$. This can be considered as the FG controlled mode. In short, $I_d$-$V_{BG}$ is the BG controlled mode when the channel conductance increases with increasing $V_{BG}$, while $I_d$-$V_{BG}$ is the FG controlled mode when the channel conductance decreases with decreasing $V_{BG}$ in this device. As seen above, this analysis on the $V_{FG}$ trajectory provides a detailed understanding of the operation mechanism behind the $I_d$-$V_{BG}$ curve.

In NVM devices, the number of charges stored in the FG at the zero bias condition ($V_{BG}$ = 0 V) is the most important parameter to guarantee their non-volatility. Therefore, both $V_{th}$ and the memory window should be determined by the number of charges stored in the FG. In case of the $V_{th}$ measurement after P/E operation, when $V_{BG}$ is swept from 0 V, the charges stored in the FG screen the electric field from the BG. Therefore, additional $V_{BG}$ is required to modulate the channel, resulting in the $V_{th}$ shift. That is, the amount of $V_{th}$ shift is indeed determined by the number of charges stored in the FG. On the other hand, in case of the $I_d$-$V_{BG}$ round sweep, the $V_{th}$ shift is not solely determined by the number of charges stored in the FG, as explained below. The situation of $V_{BG}$ switching back from positive to negative sweep at $V_{BG}$ = +30 V is considered. During the positive $V_{BG}$ sweep toward $V_{BG}$ = +30 V, $V_{FG}$ has been already fixed due to the feedback mechanism between capacitive coupling and tunneling, as shown in **Figure 3a**. However, just at the switch back to the negative $V_{BG}$ sweep, tunneling is terminated and then, $V_{FG}$ starts to decrease due to capacitive coupling. That is, the $V_{th}$ shift in the $I_d$-$V_{BG}$ round sweep is determined not only by the number of charges stored in the FG but also by capacitive coupling from the maximum $V_{BG}$ reached before the switch back. Therefore, as seen in **Figure S7**, as the sweep range of $V_{BG}$ is extended, the $V_{th}$ shift is apparently enlarged. Based on this discussion, it can be concluded that the memory window determined by the $I_d$-$V_{BG}$ round sweep is overestimated. Note that, in case of the device with no access region[22], the situation of the overestimation is same as our case. In that case, the BG controlled mode is missing and whole channel conductance is modulated by FG, since whole channel region is overlapped with the FG. Therefore, just at the switch back of the $V_{BG}$ sweep, $V_{FG}$ starts to decrease due to capacitive coupling and it forces the channel to be turned off.

Finally, the memory windows determined by the aforementioned two methods are compared. The $V_{FG}$ trajectories in the $I_d$-$V_{BG}$ round sweep with various $V_{BG}$ sweep ranges (±20, 25, 30 V) are shown in **Figure S7**. For all cases, the BG controlled region and FG controlled region were clearly observed. **Figure 4a** compares the memory windows determined with various $V_{BG}$ sweep ranges by the two methods. The horizontal axis represents the maximum $V_{BG}$ in the $I_d$-$V_{BG}$ round sweep (for filled triangles), while it represents the absolute values of the P/E voltages in the single sweep (for filled circles). Previous studies[18-20] have reported quite large memory windows and linear relationship between memory window and maximum $V_{BG}$. Although the same situation is observed in our results when the round sweep is used to determine the memory window, the overestimation of memory window is clearly shown in all three voltage conditions. Of course, since FG electrode is not compulsory for memory devices, equation (1) is not always satisfied from the geometrical viewpoint. Here, the condition that the overestimation is observed is discussed from the viewpoint of the capacitive coupling ratio. The case studies for overestimation case and consistent case, and its generalized cases are schematically illustrated in **Figure S8**. When considering whether the overestimation occurs or not, the $V_{FG}$ trajectory in the single sweep after P/E operation should be compared with that in the round sweep. However, the $V_{FG}$ trajectory in the single sweep after P/E operation cannot be measured because FG is discharged during changing the measurement function from P/E operation to $V_{th}$ measurement manually. Nevertheless, the expected trajectories can be illustrated in **Figure S8**. These figures clearly explain that the overestimation should be observed when the $V_{FG}$ trajectory of the round sweep does not match those of single sweeps. In short, the overestimation should be observed when the following equation is satisfied in the round sweep,

$$\frac{C_{ox}}{C_{ox} + C_{BN}} \cdot V_{BGmax} > V_{tun}^{(+)} - V_{tun}^{(-)}, \cdots (2)$$

where $V_{BGmax} > 0$ V is the maximum voltage applied during round sweep, and $V_{tun}^{(+)}$ and $V_{tun}^{(-)}$ are the tunnel start voltages of positive and negative sides, respectively. In case that the equation (2) is satisfied, the tunneling does occur during the $V_{BG}$ forward sweep from -$V_{BGmax}$ to 0 V. It means that some of holes stored initially at -$V_{BGmax}$ are lost until the $V_{BG}$ reaches 0 V since the number of holes stored initially at -$V_{BGmax}$ exceeds the capacity of FG at $V_{BG} = 0$ V. That is, the memory window observed in the round sweep is apparently raised by the apparent number of charges stored at -$V_{BGmax}$. The more detailed explanation can be found in **Figure S8**. Note that the equation (2) is easily satisfied by the typical geometrical and experimental situation without the FG electrode. Moreover, the memory window overestimation was also observed in $WSe_2$ and $MoS_2$ channel devices, as shown in **Figures 4b** and **4c**, respectively. These devices are fabricated by the same procedure used to fabricate the $MoTe_2$ channel device. The FG and BG controlled modes are also evident. These results suggest that the memory window overestimation is not special for $MoTe_2$ but common for all 2D channel materials as long as the $I_d$-$V_{BG}$ round sweep is used.

In conclusion, the present finding on the overestimation of memory window claims that the memory window of 2D NVM devices should be determined as same as Si NVM devices, since it is a fundamental figure of merit, and the advantage of 2D NVM devices should be fairly clarified in the same context of conventional Si NVM technologies. Moreover, the $V_{FG}$ measurements proposed in this study clearly separates the two different operation mechanisms behind the $I_d$-$V_{BG}$ curve, that is, capacitive coupling and the tunneling. This method can be used to clarify the operation mechanism in the future analysis of the various 2D materials based NMV devices.

**Experimental Section**

*Device fabrication*: Natural *n*-$MoS_2$ crystals were purchased from SPI Supplies, whereas other 2D bulk crystals were grown using a physical vapor transport technique with an $I_2$ transport agent (2H-$MoTe_2$),[33] a physical vapor transport technique without an $I_2$ transport agent ($WSe_2$)[33] and a temperature-gradient method under a high-pressure and high-temperature atmosphere (*h*-BN).[34] All 2D flakes were mechanically exfoliated from each bulk crystal and transferred onto a polydimethylsiloxane (PDMS) sheet. By using a dry transfer system with aluminum slope,[31,32] 2D flakes were sequentially transferred onto 90 nm $SiO_2/n^+$-Si substrate from

PDMS. Then, standard electron beam patterning was conducted for the source, drain and FG electrodes. Ni/Au was selected for the metal electrodes, because an ambipolar behavior of $MoTe_2$ FETs has been reported with Ni electrodes.[6] For $WSe_2$ and $MoS_2$ channels, Ni/Au electrodes were also used.

*Characterization*: Raman spectroscopy and AFM were employed to determine the crystal quality and thickness of the flakes. All the electrical measurements performed in this study were conducted in a vacuum prober with a cryogenic system by utilizing a Keysight B1500 semiconductor parameter analyzer.

**Supporting Information**

Supporting Information is available from the Wiley Online Library or from the author.

**Acknowledgements**

This research was supported by Kioxia Cooperation, the Canon Foundation, the Elemental Strategy Initiative conducted by the MEXT, Grant Number JPMXP0112101001, the JSPS Core-to-Core Program, A. Advanced Research Networks, the JSPS A3 Foresight Program, JSPS KAKENHI Grant Numbers JP20H00354, JP19H00755, 19K21956, and 18H03864, and CREST (grant number: JPMJCR15F3) commissioned by the Japan Science and Technology Agency (JST), Japan.

Received: ((will be filled in by the editorial staff))
Revised: ((will be filled in by the editorial staff))
Published online: ((will be filled in by the editorial staff))

**FIGURES**

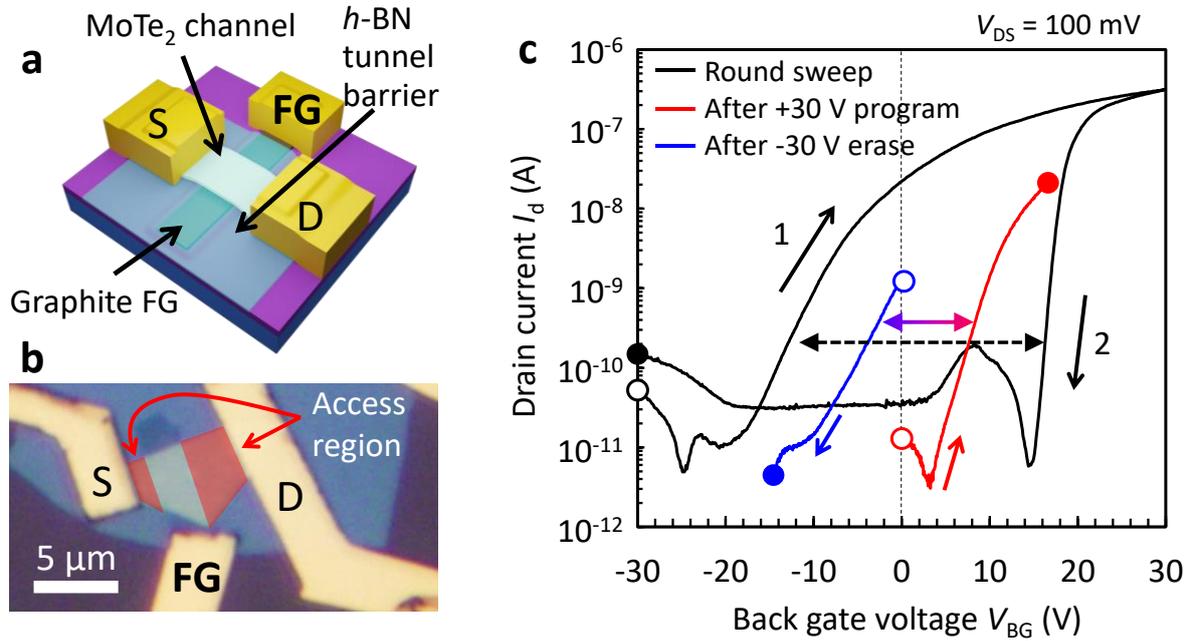

**Figure 1.** (a) Schematic and (b) optical microscope images of a fabricated 2D materials based memory device. The device consists of a 2H-MoTe$_2$ channel (8.3 nm), an $h$-BN tunnel barrier (15 nm) and a graphite FG (6.9 nm) on a 90 nm SiO$_2$/$n^+$-Si substrate. (c) $I_d$-$V_{BG}$ round sweep curves (black), showing a large memory window (black dotted arrow), and two sets of $I_d$-$V_{BG}$ single sweep curves (blue and red) after P/E operation, showing a small memory window (color arrow). Open and filled circles represent the start and end points of the $V_{BG}$ sweep, respectively.

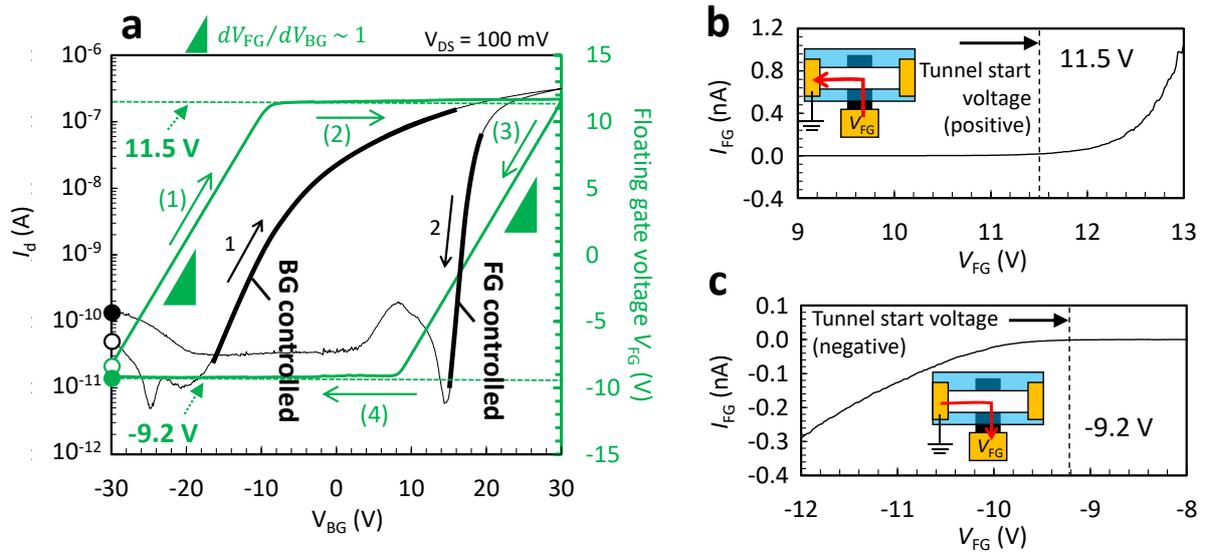

**Figure 2.** (a) Measured $V_{FG}$ trajectory in round sweep $I_d$-$V_{BG}$. Open and filled circles represent the start and end points of the sweep, respectively. (b) Tunneling current from the FG to the channel and (c) tunneling current from the channel to the FG. In the measurement, the source was grounded, the drain was floating, and $V_{BG} = 0$ V was applied.

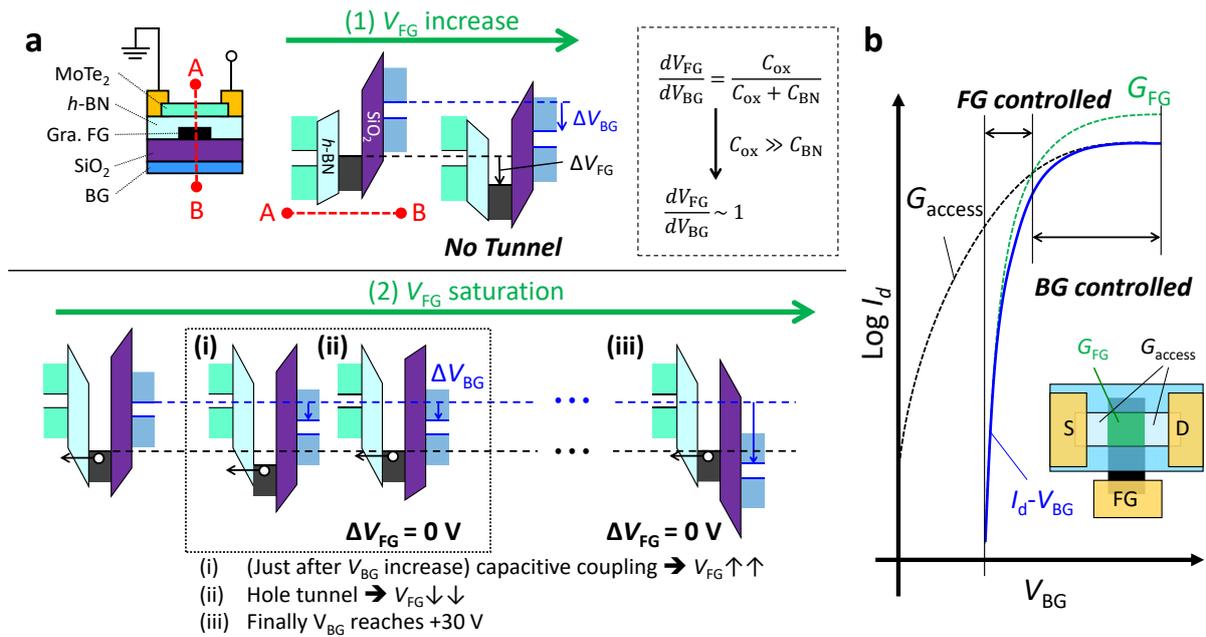

**Figure 3.** (a) Band diagrams corresponding to the $V_{FG}$ increase at the initial stage of positive sweep and following $V_{FG}$ saturation. The stages of (1) and (2) correspond to those of (1) and (2) with the green color in **Figure 2a**. (i) and (ii) indicate the feedback mechanism between capacitive coupling and hole tunneling at stage (2) in the positive $V_{BG}$ sweep. (iii) Band diagram just at $V_{BG}$ = +30 V, where $V_{FG}$ is still fixed at the tunnel start voltage. (b) Schematic showing the BG controlled mode and FG controlled mode in the $I_d$-$V_{BG}$ curve. The conductance between the source and drain is determined by the smaller of $G_{access}$ and $G_{FG}$.

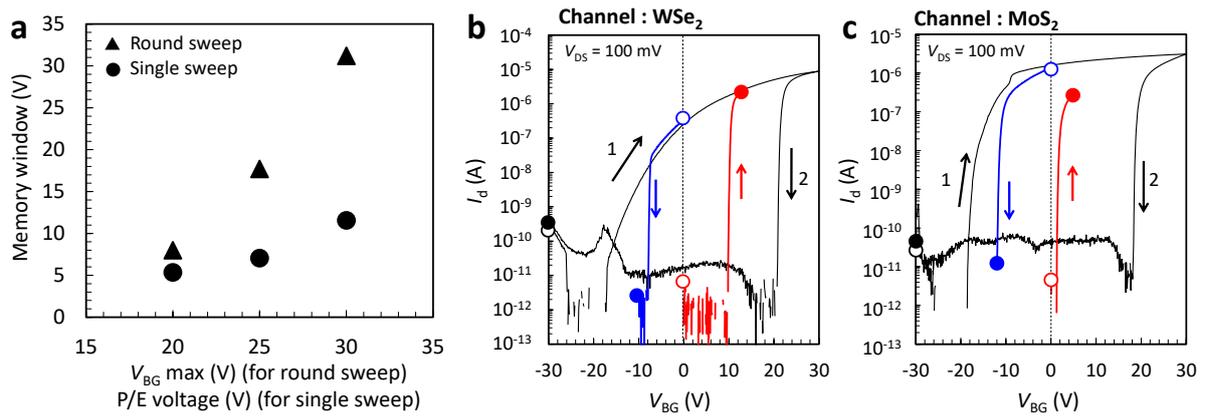

**Figure 4.** (a) Comparison of the memory windows obtained by two kinds of methods for different voltage conditions. (b), (c) $I_d$-$V_{BG}$ round sweep curves (black) and $I_d$-$V_{BG}$ single sweep curves from 0 V after the +30 V program (red) and after -30 V erase (blue) of the WSe$_2$ and MoS$_2$ channel devices. Blank and filled circles represent the start and end point of the $V_{BG}$ sweep, respectively. Memory window overestimation was also observed in these two devices.

The table of contents entry

The floating gate voltage measurement, which is proposed here, reveals the memory window overestimation of floating gate type 2D memory devices. Although round sweep transfer curves are usually used to determine the memory window in 2D memory research field, it is not good choice. The memory window should be determined in the same way of Si memory research field. (59 words)

*Keyword 2D materials, 2D hetero-stack, non-volatile memory, floating gate, floating gate voltage*

*Taro Sasaki, Keiji Ueno, Takashi Taniguchi, Kenji Watanabe, Tomonori Nishimura, and Kosuke Nagashio\**

**Title** Understanding the Memory Window Overestimation of 2D Materials Based Floating Gate Type Memory Devices by Measuring Floating Gate Voltage

*ToC figure*

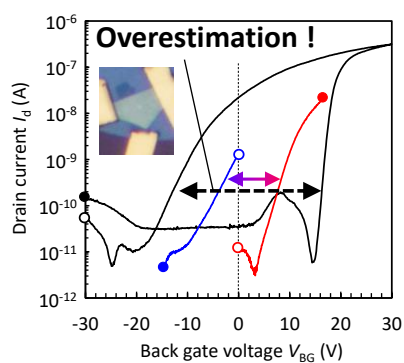

## Supporting Information

**Understanding the Memory Window Overestimation of 2D Materials Based Floating Gate Type Memory Devices by Measuring Floating Gate Voltage**

*Taro Sasaki, Keiji Ueno, Takashi Taniguchi, Kenji Watanabe, Tomonori Nishimura, and Kosuke Nagashio\**

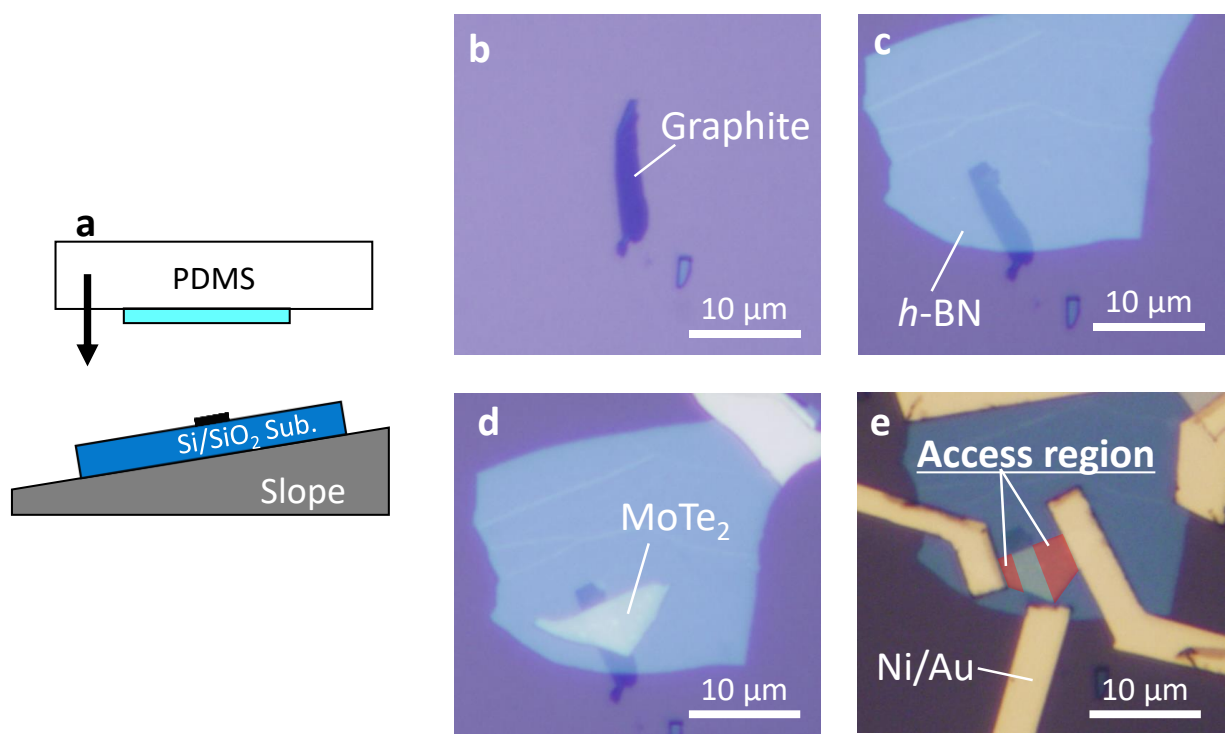

**Figure S1**. Details of the device fabrication procedure. (a) Schematic of the 2D stacking process. A mechanically exfoliated 2D flake is pre-transferred onto PDMS and subsequently stamped to the target substrate or the other 2D flake. An aluminum slope was used to suppress the bubbles between the 2D flakes. Optical images of (b) after graphite transfer on the $SiO_2$/Si substrate, (c) after $h$-BN transfer on graphite, (d) after $MoTe_2$ transfer on $h$-BN/graphite and (e) after Ni/Au metal electrode deposition. The access region is highlighted in (e).

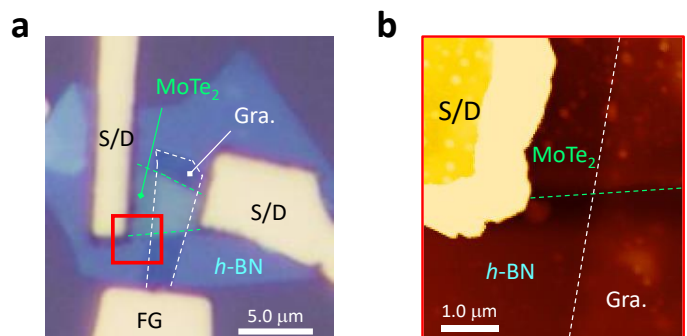

**Figure S2.** (a) Optical microscope image and (b) AFM image of typical MoTe₂/*h*-BN/Graphite hetero-stack based memory device. The AFM image was taken at the red square region in (a).

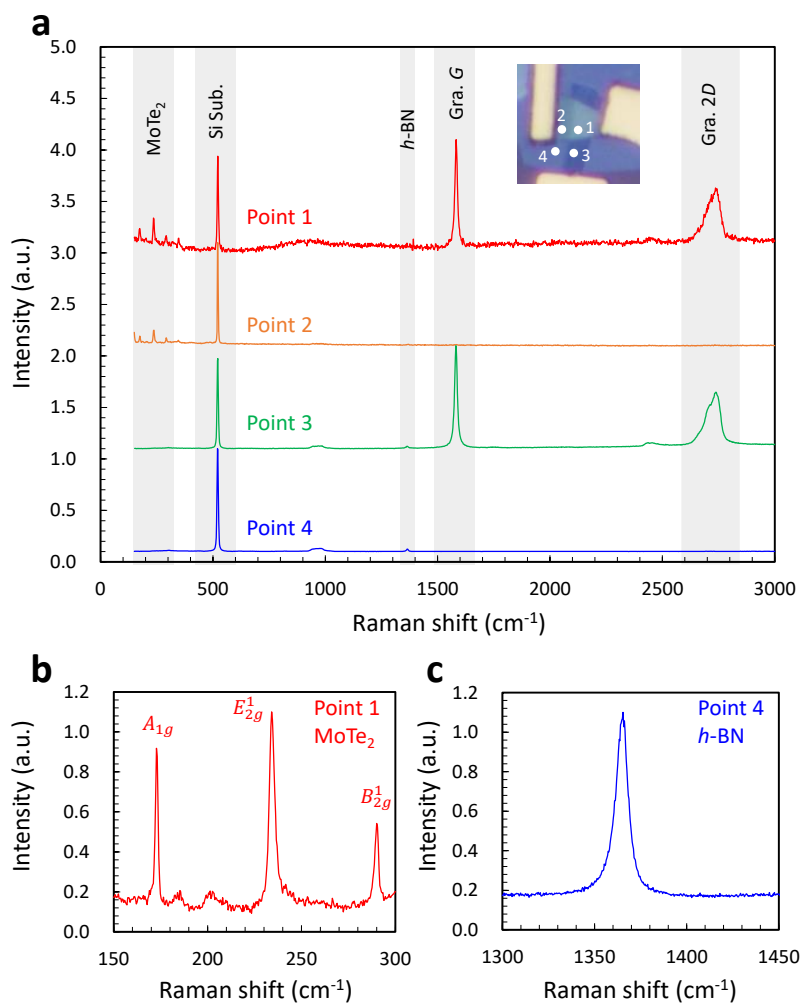

**Figure S3.** Raman spectra of typical MoTe$_2$/*h*-BN/Graphite hetero-stack based memory device. (a) Raman spectra with wide range of Raman shift at four points indicated in the inset. Point 1 corresponds to the MoTe$_2$ (5.87 nm)/*h*-BN (9.46 nm)/Graphite (8.73 nm) stack. Similarly, points 2, 3 and 4 correspond to the MoTe$_2$/*h*-BN stack, *h*-BN/Graphite stack and *h*-BN, respectively. Each spectrum well agrees with those of the stacking materials. The *h*-BN peak is not observed at points 1 and 2 since MoTe$_2$ is overlapped on *h*-BN. (b) Enlarged MoTe$_2$ spectra of point 1 and (c) enlarged *h*-BN spectra of point 4 well agree with the previous Raman studies [1,2]. Laser powers are 0.06 mW for points 1 and 2, and 0.38 mW for points 3 and 4. Laser spot size and wavelength are ~1 μm and 488 nm, respectively.

Reference. [1] M. Grzeszczyk *et. al.*, *2D Mater.*, **2016**, *3*, 025010. [2] R. V. Gorbachev *et. al.*, *small*, **2011**, *7*, 465.

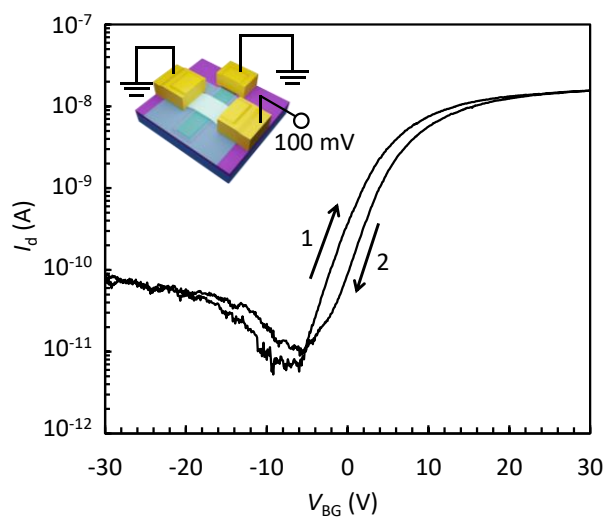

**Figure S4.** $I_d$-$V_{BG}$ round sweep transfer curves with the grounded FG. Inset shows the bias condition. When the FG was grounded during the round $V_{BG}$ sweep, the large memory window observed in **Figure 1c** disappeared, and quite a small hysteresis was observed, because the graphite FG acts as a local back gate.

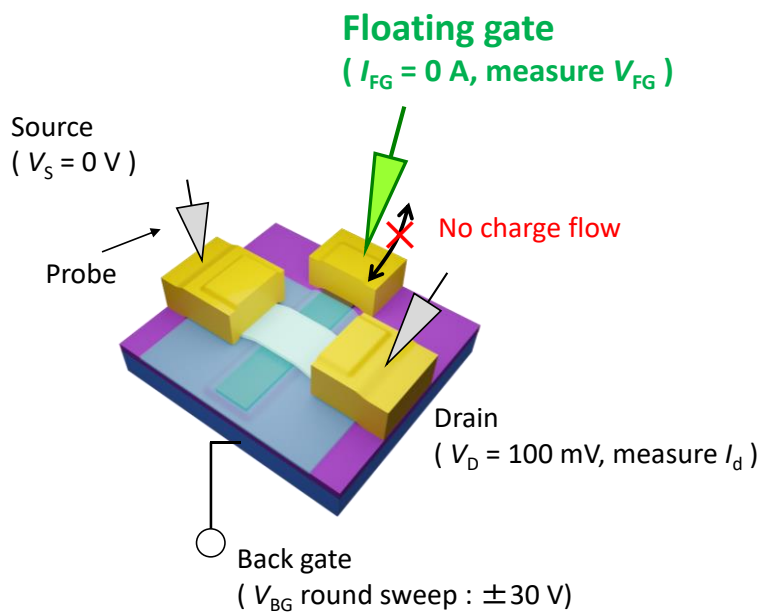

**Figure S5.** The measurement condition used to monitor $V_{FG}$. Zero current condition applied to FG electrode prevents the charge loss from the FG to the electrode, which enables the measurement of $V_{FG}$.

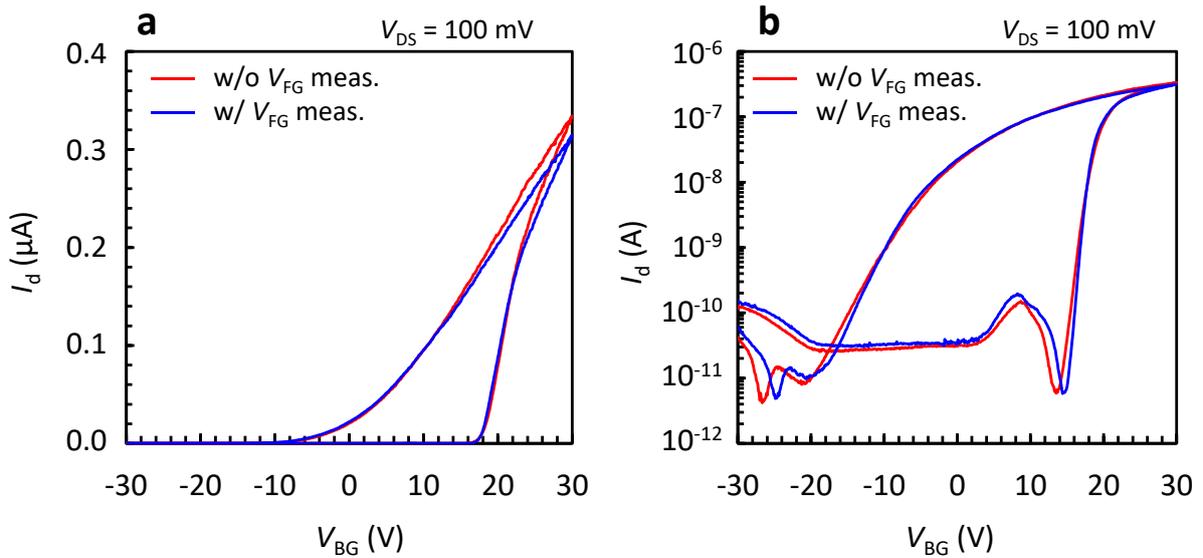

**Figure S6.** Round sweep $I_d$-$V_{BG}$ measurement with and without the $V_{FG}$ measurement. (a) Linear scale and (b) semi-log scale. It has been confirmed that the $V_{FG}$ measurement itself has a negligible effect on the $I_d$-$V_{BG}$ measurement.

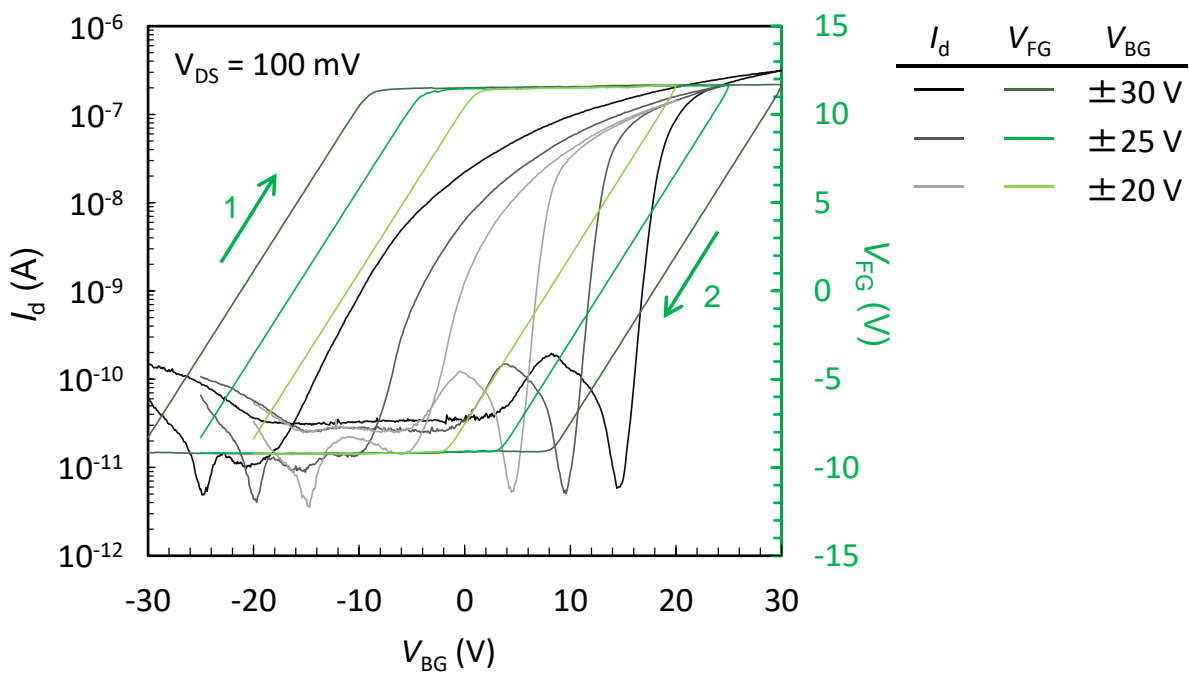

**Figure S7.** $V_{FG}$ trajectories superimposed over $I_d$-$V_{BG}$ curves with various $V_{BG}$ ranges.

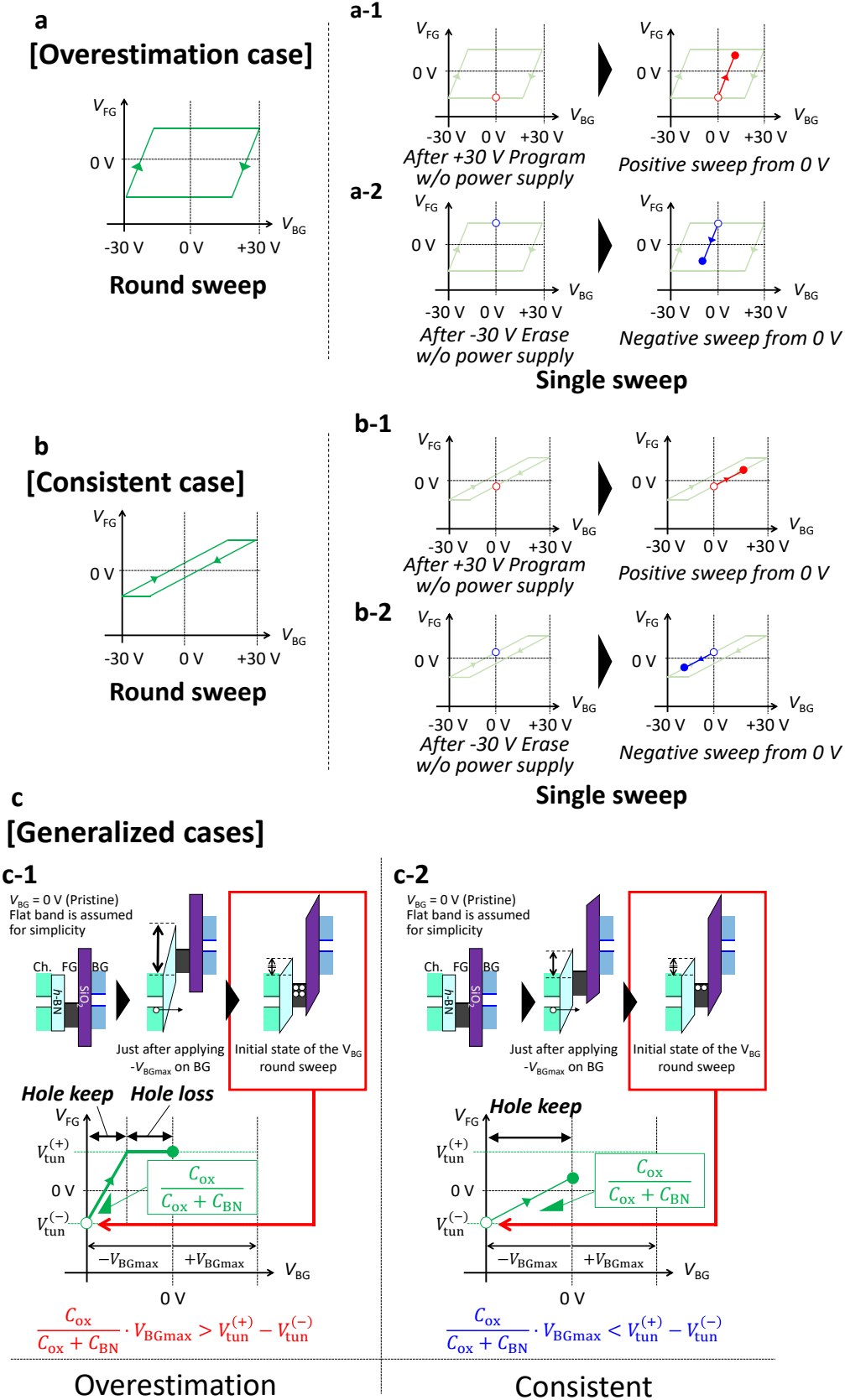

**Figure S8.** The case study for (a) overestimation case (high capacitive coupling ratio) and (b) consistent case (low capacitive coupling ratio), and (c) its generalized cases. Expected $V_{FG}$ values after P/E operation without power supply ($V_{BG} = 0$ V) are illustrated as red and blue blank circles, and expected $V_{FG}$ trajectories for the single sweep are illustrated as red and blue line in (a) and (b).

**Note:** Although the reduction of capacitive coupling ratio ($C_{ox} / C_{ox} + C_{BN}$) leads to a negligible overestimation, the overestimation is not solely determined from the ratio. Important point is that the overestimation should be observed when $V_{FG}$ trajectory of round sweep does not match with those of single sweeps (The reason is discussed below).

The case study and its generalized cases are illustrated in **Fig. S8.** Unfortunately, the $V_{FG}$ trajectory in single sweep after P/E operation can't be measured because FG is discharged during changing the measurement function from P/E operation to $V_{th}$ measurement manually. However, the expected $V_{FG}$ trajectories for single sweep after P/E operation can be illustrated. **Fig. S8a** shows typical overestimation case with the capacitive coupling ratio of 1 observed during the round sweep. Here, for the single sweep case, the expected $V_{FG}$ value after program operation without power supply and its trajectory during positive sweep from 0 V are indicated as the red open circle and red line in **a-1**, respectively. $V_{FG}$ is linearly increased with $V_{BG}$ due to capacitive coupling. Similarly, the expected $V_{FG}$ trajectory of single sweep after erase operation is shown in **a-2**. As you can see, $V_{FG}$ trajectory of round sweep and those of single sweeps do not match. Therefore, an overestimation is observed. On the other hand, consistent case with low capacitive coupling ratio is shown in **Fig. S8b**. The expected $V_{FG}$ trajectory after program operation without power supply follows the round sweep $V_{FG}$ trajectory as shown in **b-1** (vice versa in **b-2**). In this case, both trajectories match, and memory window determined from round sweep is expected to be small.

Above mentioned discussion can be generalized as shown in **Fig. S8c.** When the overestimation is observed "in the round sweep", following equation should be satisfied.

$$\frac{C_{ox}}{C_{ox} + C_{BN}} \cdot V_{BGmax} > V_{tun}^{(+)} - V_{tun}^{(-)}, \cdots (2)$$

where $V_{BGmax}$ ($> 0$ V) is the maximum voltage applied during $V_{BG}$ round sweep, $V_{tun}^{(+)}$ and $V_{tun}^{(-)}$ are the tunnel start voltages of positive and negative sides, respectively. Now let's discuss the validity of this criteria. First, we consider the case that the equation (2) is satisfied, as shown in **c-1**. When $V_{BG} = -V_{BGmax}$ is initially applied, many holes are stored before the round sweep (for example, 100 holes), as shown in the band diagram in **c-1**. Then, $V_{BG}$ is increased to 0 V. The tunneling does occur at the tunnel start voltage of $V_{tun}^{(+)}$ until $V_{BG}$ reaches 0 V. In this case, some of stored holes should be lost by the tunneling (for example, 70 holes are kept at $V_{BG} = 0$ V, but 30 holes are lost.). Under this condition, when single sweep from $V_{BG} = 0$ V to negative side is carried out, $V_{FG}$ follows the blue line in **a-2**. That is, $I_d$-$V_{BG}$ for the single sweep corresponds to the "70 holes case", even though the forward sweep of $I_d$-$V_{BG}$ in the round sweep apparently corresponds to the "100 holes case". On the other hand, the case that the equation (2) is not satisfied is considered (low capacitive coupling ratio case). Since capacitive coupling ratio is low, the voltage across $h$-BN just after applying $-V_{BGmax}$ on BG (before feedback by tunneling) is less than that of high capacitive coupling ratio case shown in **c-1**. Therefore, the number of holes stored initially is also less than the case shown in **c-1** (for example, 50 holes). As shown in **c-2**, when $V_{BG}$ swept from $-V_{BGmax}$ to 0 V, tunneling does not occur during forward sweep and, finally, $V_{BG}$ reaches 0 V. In this case, all the holes are kept at $V_{BG} = 0$ V (50 holes). Under this condition, when single sweep from $V_{BG} = 0$ V to negative side is carried out, $V_{FG}$ follows the blue line in **b-2**. Both of $I_d$-$V_{BG}$ for the round sweep and single sweep correspond to "50 holes case". That is, equation (2) is the criteria or the overestimation. Therefore, the overestimation is not only determined from the capacitive coupling ratio but also sweep range and tunnel start voltages.